\documentclass[preprint,12pt]{elsarticle}

\usepackage{makecell}
\usepackage{multirow}
\usepackage{wrapfig}
\usepackage{makecell}
\usepackage{amssymb}
\usepackage{amsmath}
\usepackage{eurosym}
\usepackage{adjustbox}
\usepackage{ulem}
\usepackage{xcolor}

\newcommand{\asr}[1]{\textcolor{black}{#1}}
\newcommand{\revone}[1]{\textcolor{black}{#1}}
\newcommand{\revtwo}[1]{{\leavevmode\color{black}#1}}

\journal{Urban Climate}

\begin{document}
\begin{frontmatter}



\title{City-Scale Assessment of Pedestrian Exposure to Air Pollution: A Case Study in Barcelona}

\author[UPC,BSC]{Jan Mateu Armengol}
\author[BSC]{Cristina Carnerero}
\author[UOC]{Clément Rames}
\author[BSC]{Álvaro Criado}
\author[UOC]{Javier Borge-Holthoefer}
\author[BSC]{Albert Soret}
\author[UOC]{Albert Solé-Ribalta}

\affiliation[UPC]{organization={Department of Fluid Mechanics, Universitat Politècnica de Catalunya},
            city={Barcelona},
            postcode={08034}, 
            country={Spain}}
\affiliation[BSC]{organization={Barcelona Supercomputing Center},
            city={Barcelona},
            country={Spain}}
\affiliation[UOC]{organization={Internet Interdisciplinary Institute (IN3), Universitat Oberta de Catalunya},
            city={Barcelona},
            state={Catalonia},
            country={Spain}}

\begin{abstract}

Air pollution is a pressing environmental risk to public health, particularly in cities where population density and pollution levels are high. Traditional methods for exposure analysis often rely on census data, but recent studies highlight the impact of daily mobility on individuals' exposure. \revtwo{Here, we develop a methodology to determine unprecedented pedestrian exposure estimates at the city scale by combining sidewalk pedestrian flows with high-resolution (25 m $\times$ 25 m) NO$_2$ data from bias-corrected predictions of the air quality system CALIOPE-Urban. Applied to Barcelona (Spain) for the year 2019, we show that pedestrian flow and NO$_2$ levels exhibit negligible temporal correlation. While short-term (hourly) exposure is driven by pedestrian mobility, long-term (monthly) exposure is dominated by NO$_2$ patterns. We identify strong spatial gradients of exposure, highlighting the importance for high-resolution solutions at the sidewalks scale. Finally, we determine that exposure mitigation strategies should consider different citizen subgroups based on their mobility and preferred routes, as significant differences were found between residential and pedestrian exposure. Our results provide exposure indicators designed for city planners and policymakers, helping to prioritize mitigation measures where and when they are most needed.}

\end{abstract}



\begin{keyword}
Pedestrian exposure \sep urban air quality \sep pedestrian mobility \sep Policy and planning
\end{keyword}

\end{frontmatter}


\section{Introduction}
\label{sec:intro}

Air pollution has reached a disease burden comparable to other major health risks in our society, such as unhealthy diets and tobacco smoking, \revone{it is currently considered the most significant environmental threat to global public health according to the World Health Organization (WHO) \cite{who2021who}}. Short- and long-term exposure to air pollution have been mainly linked to adverse effects on respiratory and cardiovascular health \cite{lee2014air,guan2016impact}. Additionally, evidence suggests that it can impair cognitive development, increase the risk of type 2 diabetes, contribute to cancer development, accelerate skin aging, and act as a risk factor for obesity \cite{holgate2017every}. In Europe, during 2019, the majority of the urban population experienced concentrations levels above the WHO guidelines \cite{eea2019air}, 94\% for nitrogen dioxide ($\mbox{NO}_2$) and 97\% for particulate matter (PM). The situation in Barcelona (Spain) is alike: high outdoor levels of NO$_2$, mainly caused by road traffic \citep{casquero2019impact}, have significant adverse effects on both public health and the economy. A recent study \cite{font2023estimating} revealed that the estimated total mortality in the city attributable to PM2.5 and NO$_2$ air pollution stands at 13\%, with an associated annual social cost of 1,292 million euros.

Due to the daily mobility and behavior of city dwellers, commuting is considered one of the \revtwo{periods} of highest exposure to air pollution, particularly because of the proximity to vehicle traffic \cite{de2012travel,karanasiou2014assessment}. Disregarding the analysis by transport mode, the exposure of car drivers to air pollution is similar to or higher than that of pedestrians, motorcyclists, or cyclists. However, car ventilation systems can significantly reduce the inhaled dose of pollutants for vehicle occupants, particularly for particulate matter  \cite{cepeda2017levels}. In contrast, pedestrians can generally only reduce their exposure by altering their mobility patterns, such as distancing themselves from heavily trafficked streets, choosing routes with green spaces, or avoiding travel during rush hours. Thus, the appropriate characterization of pedestrian exposure, which needs to rely on high-resolution estimations of pollutant concentrations and detailed sidewalk mobility demand, plays a vital role in these epidemiological studies and it is essential for air quality authorities striving to mitigate the health impacts of air pollution.

Considering these population health concerns, traditional epidemiological studies estimate pollutant concentration levels based on government monitoring data \cite{pope2009fine,miller2007long,laden2006reduction,chuang2007effect}. This approach implicitly assumes that monitoring stations accurately represent large areas and that all residents within a region experience equivalent levels of exposure. Although this method provides valuable information, the typically limited number of monitoring stations challenges their spatial representativeness. Some of these limitations can be addressed by using models to enhance spatial resolution at residential addresses and to capture the highly heterogeneous spatial distribution of pollution in urban environments \cite{VARDOULAKIS20052725, SANTIAGO201361, DUYZER201588}. These models include Land Use Regression  \cite{kramer2009eczema,de2014airway}, mesoscale air quality \cite{boldo2014air,izquierdo2020health}, Gaussian dispersion \cite{willers2013fine,batterman2014comparison,korek2015traffic,van2016health}, and Computational Fluid Dynamic (CFD) \cite{rivas2019cfd}. However, this approach, commonly referred to as \textit{static exposure estimation}, still assumes that the population remains in the same location throughout the day, an assumption that is often too rigid.

As a counterpart, several studies have emphasized the importance of considering human mobility \cite{park2017individual,reis2018influence,dias2018spatial}. Since pollution concentrations vary across different city areas, incorporating this dynamic information can significantly improve the estimations. The work of Nyhan \textit{et al.} \cite{nyhan2016exposure} found that PM2.5 exposure significantly differs ($p < 0.05$) for most districts of New York City when citizen mobility is considered, compared to \textit{static exposure} approach. However, implementing this approach in practice presents significant challenges for the transport research community \revone{due to data scarcity and the complexity of the transportation network structures}. Traditional methods for estimating pedestrian flow have primarily centered on agent-based models \cite{batty2001agent}, with recent advancements leveraging specialized micro-simulation software such as Viswalk (part of VISSIM \cite{fellendorf2010microscopic}) and MoPeD \cite{zhang2022assessing}. These tools may provide highly accurate mobility insights but at the same time require significant setup efforts and precise data, limiting their application to small areas \cite{zhang2022assessing,santiago2021estimates,puusepp2018simulating}. While several works exploiting Machine Learning techniques are starting to appear \cite{hankey2021predicting,jiang2022pedestrian}, contemporary strategies leverage network theory and complex systems \cite{ye2016modified,sevtsuk2021estimating}. In our opinion, these latter approaches achieve a notable balance between scalability and precision in results.

\revone{Two distinct approaches exist to estimate pollution exposure} depending on the research question: conducting the evaluation based on citizen subgroups or specific geographic areas. \textit{Population-focused exposure}, central to epidemiological studies, characterizes the accumulated dose in a particular group or subgroup of individuals and seeks to infer causal correlations between air pollution and public health. However, these metrics do not provide direct actionable insights related to the spatial layout of cities, as \textit{area-focused exposure} indicators do. In particular, area-focused analysis aims to estimate the extent to which the distribution of pollutants impacts the (transiting through) population in specific areas. Such spatial analysis is therefore fundamental for urban planners and decision-makers to prioritize mitigation strategies where they are most needed, as they identify exposure hot-spots (i.e., crowded areas suffering from severe air pollution). 

Despite their informative potential, only a limited number of scholars have addressed \textit{area-focused exposure} indicators in urban settings at various resolution scales. Following the pioneering work of Nyhan \textit{et al.} \cite{nyhan2016exposure} in New York City, which relied on interpolated NO$_2$ data at different district centroids, \revone{Picornell \textit{et al.} \cite{picornell2019population} reported \textit{area-focused} exposure indicators for NO$_2$ using a mesoscale air quality model at a 1 km $\times$ 1 km spatial resolution. Despite considering the pollution spatio-temporal variability,} mesoscale spatial resolution typically cannot capture the strong NO$_2$ gradients found in urban areas \cite{criado2023data}. To overcome these precision challenges, Santiago \textit{et al.} \cite{santiago2021estimates} conducted a high-resolution \textit{area-focused exposure} study by combining a detailed microscale pedestrian model \cite{fellendorf2010microscopic} with NO$_2$ data from Computational Fluid Dynamic simulations. Their 19-day case study covered a spatial domain of 300 m $\times$ 300 m around \textit{Plaza Elíptica} square in Madrid. Despite the computational demand of the model which restricts the modeling area, they showed that some exposure peaks, especially at bus stops and crosswalks, cannot be detected at larger spatial scales. 


In this context, our work proposes a pipeline to provide unprecedented high-resolution estimates of pedestrian NO$_2$ \revone{\textit{area-focused exposure}. Pedestrian flows} on each sidewalk segment of the city are estimated by adjusting a mathematical model using data from 147 pedestrian count sensors with hourly resolution deployed across Barcelona in 2019. On the other hand, hourly NO$_2$ outdoor concentrations are estimated using the bias-corrected urban air quality model CALIOPE-Urban \cite{criado2023data}. The exposure model we propose is applied across Barcelona using data from 2019 and allows us to achieve three distinct objectives. First, we reveal the temporal profiles of NO$_2$ concentration and pedestrian flows, which interestingly do not exhibit strong temporal and spatial correlation, and then combine them to determine pedestrian exposure. Second, we spatially identify local exposure hot-spots in the city and conclude that accurately determining these areas requires high-resolution data (or estimates) on both pedestrian flow and pollutant concentrations. Third, we compare the long-term spatial patterns of residential versus pedestrian exposure affecting Barcelona city and evidence their main spatial differences. Building upon these results, we demonstrate the effectiveness of our methodology in providing decision-makers with critical insights to prioritize urban policies aimed at reducing pedestrian $\mbox{NO}_2$ exposure.

\section{Data and Methods}

\subsection{Pedestrian mobility estimates}
\label{sec:ped_estimates}

\begin{figure}[h!]
    \includegraphics[width=0.9\textwidth]{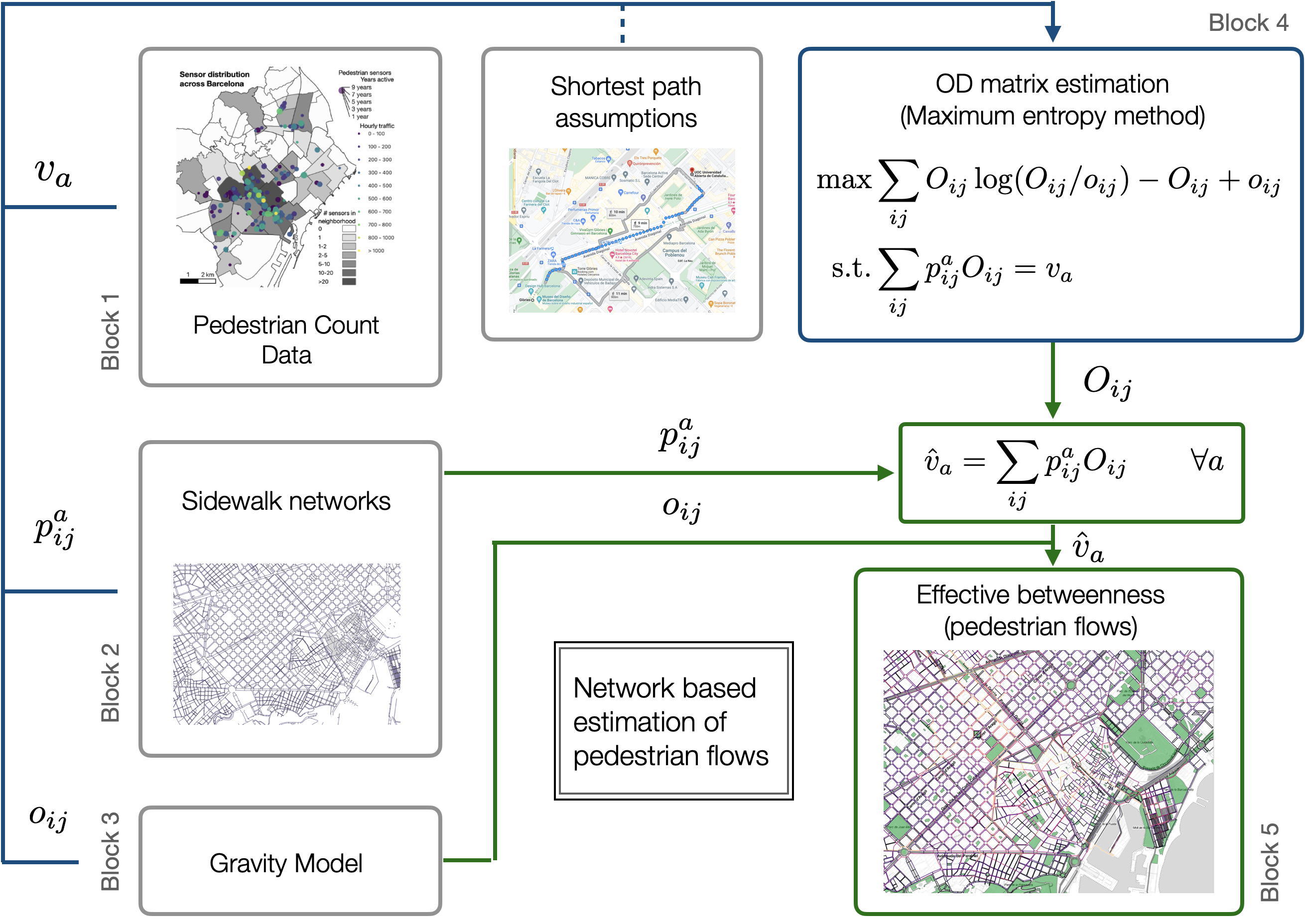}
    \centering
    \caption{Pipeline for pedestrian mobility demand. Blue lines represent the part of the pipeline dedicated to estimating the Origin-Destination (OD) matrix for pedestrian mobility. Green lines denote the part focused on estimating pedestrian flows, which occurs after the OD matrix has been estimated. Variable $v_a$ stands for the pedestrian counts obtained with the light-interrupted count sensors. Variable $p_{ij}^a$ stands for the fraction of shortest paths that pass by node $a$ when considering a route from $i$ to $j$. Variable $o_{ij}$ stands for an initial estimation of the OD-matrix, that will be used as a prior for the optimization. Variable $O_{ij}$ represents the estimated OD-matrix when all previous information is considered. Finally, variable $\hat{v}_a$ denotes the estimated pedestrian flow, which, when combined with the OD matrix, provides the final estimation.}
    \label{fig:ped_estim_pipeline}
\end{figure}

The proposed pipeline used to estimate pedestrian flows at sidewalk scale is fully outlined in Fig.~\ref{fig:ped_estim_pipeline}. Below, we detail the objective and functioning of each component.

\subsubsection{Pedestrian count data and sidewalk network infrastructure}

\begin{figure*}[t]
	\begin{tabular}{l l}		
        (a) & (b)\\	
        \includegraphics[width=0.49\textwidth]{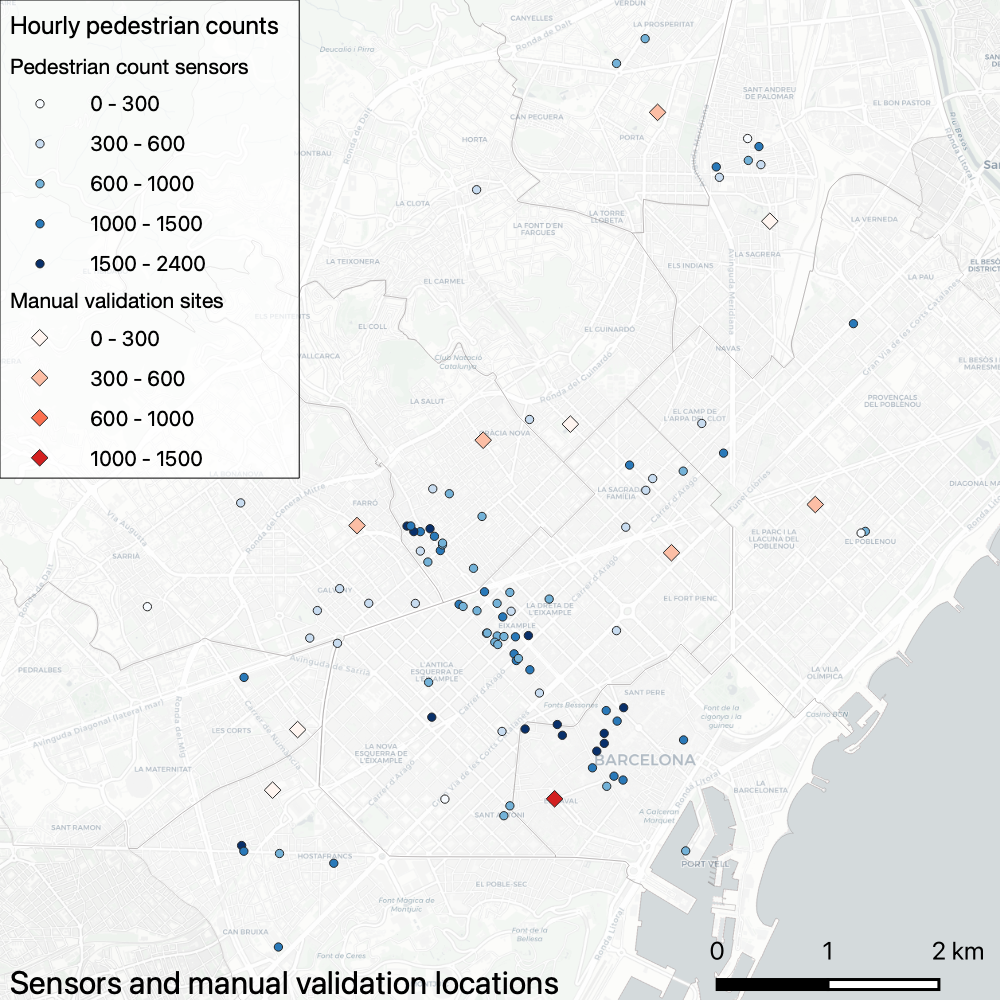} &
    	\includegraphics[width=0.465\textwidth]{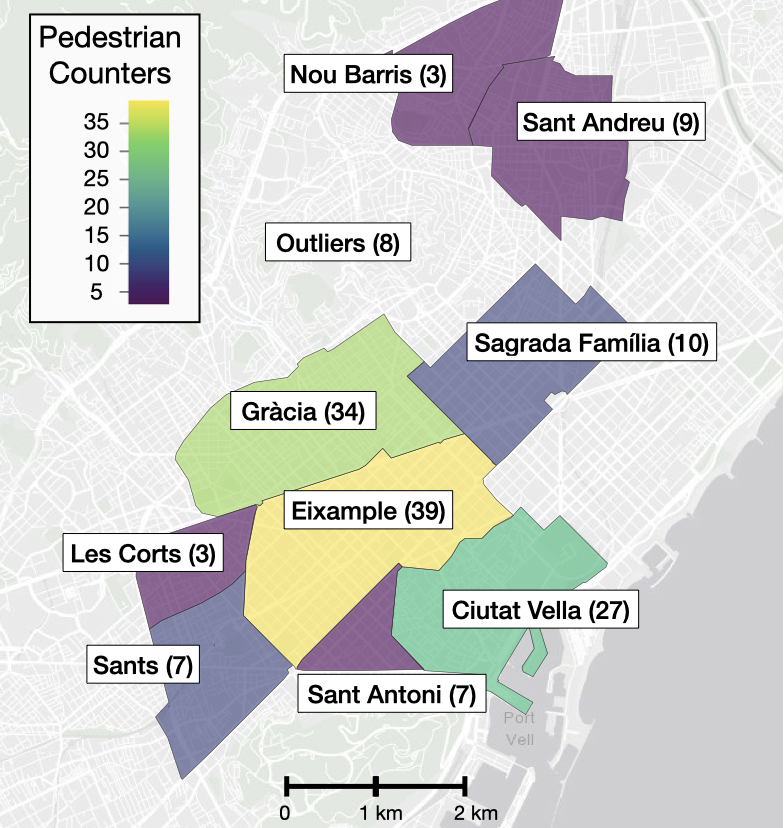}
    \end{tabular}
    \caption{(a) Locations of count sensors (with added random perturbations for anonymity) and manual validation sites. (b) Resulting clusters of pedestrian sensors after applying the  DBSCAN clustering algorithm. }
    \label{fig:manual_validation}
\end{figure*}

Obtaining reliable mobility data is a critical step to accurately calibrate pedestrian mobility models. Although a range of data sources have been used in the literature for directly monitoring pedestrian traffic in urban environments (e.g., mobile phone and GPS traces), our methodology specifically employs footfall data obtained from light-interrupted count sensors supplied by TC Group Solutions \cite{TCstreet}. Fig.~\ref{fig:manual_validation} displays the geographical distribution of these sensors. \revone{In this study, we were unable to select the sensor locations, as they are set by the customers of TC Group Solutions.} These sensors can count individual pedestrians separated by at least 25 cm within a 3 to 5 meter range from the building facades. \asr{The characteristics and locations of the sensors posed three main challenges for data collection: limited sidewalk coverage (as some sidewalks are wider than 5 meters), potential undercounting of pedestrians walking in groups, and overcounting near bar terraces. To address these issues, we applied several adjustments to the sensor counts. First, we linearly scaled the counts based on the sensor’s range (determined by its installation angle) and the sidewalk width, assuming a homogeneous distribution of pedestrians. Second, we adjusted the counts to account for the expected frequency of pedestrians walking in groups, following the findings of Harms et al. \cite{harms2019walking}. Finally, we visually inspected the sensor locations and excluded those with bar terraces adjacent. The adjusted counts were validated through fieldwork, where manual counts were compared to the sensor data both before and after scaling. The results showed a improvement in correlation, particularly for sensors with higher pedestrian flow volumes.} The resulting dataset consists of 147 sensors providing aggregated counts at hourly resolution throughout 2019. Occasionally, multiple sensors may sample the same sidewalk edge. In these situations, pedestrian flows sampled by the sensors are averaged and associated to that sidewalk segment $a$ at time $t$, $v_a(t)$.

\asr{Panel (a) of Fig. \ref{fig:manual_validation} shows the geographical distribution of the sensors across Barcelona, with some random noise added for privacy reasons. To analyze exposure at the area scale directly from the sensors, we applied DBSCAN clustering \cite{ester1996density}, using a bandwidth of 500 meters and a minimum cluster size of 3 sensors. This resulted in the formation of nine clusters, as displayed in Panel (b) of Fig. \ref{fig:manual_validation}.} Clusters are labeled with the name of the district or neighborhood they span, for the sake of interpretability.

\asr{Regarding the pedestrian infrastructure of Barcelona, we based our analysis on the sidewalk data used in the study by Rhoads et al. \cite{rhoads2021sustainable}. The sidewalk network is represented as a graph, where nodes correspond to intersections where pedestrians make routing decisions, and edges represent sidewalk segments connecting those intersections. After performing minor GIS corrections to address errors and optimize the network, we compiled a graph consisting of approximately 28,000 nodes and 43,000 edges. This graph includes sidewalks on both sides of the street, as well as crosswalks to facilitate pedestrian crossings. For further details, we refer the reader to the original publication for Barcelona \cite{rhoads2021sustainable}.}

\subsubsection{Pedestrian flows estimation}

Drawing upon classical methodologies in mobility demand analysis \cite{de2011modelling,willumsen1978estimation}, we have engineered a robust semi-analytical mobility model to estimate pedestrian demand and capture the most important complexities of urban mobility. The central step in the pipeline (Block 4 in Fig.~\ref{fig:ped_estim_pipeline}) entails estimating an OD matrix that is consistent with the pedestrian count \asr{given by the sensors}.

Several traditional approaches exist to solve this problem. We chose the maximum entropy model \cite{de2011modelling}, which estimates the OD matrix, $O_{ij}$, using minimal presupposed information. Drawing upon the principles of statistical mechanics, the algorithm attempts to identifying the OD-matrix that supports the highest number of meso-states compatible with our empirical observations, or the macro-state (the observed amount of pedestrian counts in each sensor). This process is typically formulated as an optimization problem aimed at finding the OD matrix that maximizes entropy given some constraints, which after several approximations and algebraic manipulations (see \cite{gomez2019impact}), becomes equivalent to minimizing the function
\begin{eqnarray} \label{eq:max_entropy}
        &&\log(W) = - \sum_{ij}O_{ij} log(O_{ij}) - O_{ij}, \\
        &&\mbox{such that} \sum_{ij} p_{ij}^a O_{ij} = v_a,
\end{eqnarray}
where $v_a$ are the pedestrian counts, as previously defined and $p_{ij}^a$ represents the amount of shortest paths that traverse node $a$ for movements from node $i$ to $j$ on the sidewalk network considered. Even though other types of pedestrian routing approaches are explored in the literature \cite{tang2018deviation,bongiorno2021vector}, we focus on the shortest path model because of its simplicity and realism. For example, Bongiorno \textit{et al.} \cite{bongiorno2021vector} indicate that pedestrians typically deviate by no more than 10\% from the shortest path. Although the optimization problem outlined in Eq.~\ref{eq:max_entropy} offers a practical framework, its accuracy depends on the sensor distribution throughout the modeling area. Consequently, it is typically advantageous to incorporate some prior information on expected mobility, in our case we use a gravity model as defined in \ref{sec:OD}. In this case, the optimization problem in eq. \ref{eq:max_entropy} can be recast as follows \cite{de2011modelling}:

\begin{equation}
\label{eq:2}
\log (W') = - \sum_{ij} O_{ij} \log \left( \frac{O_{ij}}{o_{ij}} \right) - O_{ij} + o_{ij},
\end{equation}

\noindent which may be related to the Kullback–Leibler divergence between the prior information and the OD matrix we are estimating. Similar approaches for estimating the OD matrix exist \cite{sevtsuk2021estimating}, each relying on different assumptions. We choose this approach because it does not inherently assume an equivalent scaling of trips between services of the same type, unlike the method described in \cite{sevtsuk2021estimating}.
 
Finally, once the OD matrix $O_{ij}$ has been obtained, the flows at each network node $(f_a)$ and network edge ($f_{kl}$) can be estimated using 
\begin{equation}
f_a~=~\sum_{ij} p_{ij}^a O_{ij},~\mbox{and}~f_{kl~}~=~\sum_{ij} p_{ij}^{kl} O_{ij},
\end{equation}
respectively. In this situation, $p_{ij}^{kl}$ corresponds to the specific counterpart of $p_{ij}^a$, representing the proportion of the shortest paths that pass through edge $kl$ for trips travelling from location $i$ to $j$ within the network. \revone{Note that the estimated pedestrian flows do not take into account whether pedestrians are moving in one direction or the other.}

\subsubsection{Accuracy of the model}

To assess the reliability of the proposed model, its accuracy was validated through two distinct experiments: (i) Leave-One-Out Cross-Validation (LOOCV) using sensor data and (ii) external manual pedestrian counts at 10 locations of the city where no sensors are installed. In the first experiment, we aim to validate our estimations within the dense and complex urban environment of the city center, where car traffic tends to accumulate and higher pedestrian densities are detected. To this end, we considered the set of sensors in the central districts of Barcelona (\textit{Ciutat Vella}, \textit{Eixample}, and \textit{Gràcia} clusters, see Fig.~\ref{fig:manual_validation}a). Eventually, this resulted in a total of 75 sensors considered for the LOOCV, which corresponds to 75 LOOCV experiments. \asr{In the second experiment, we expanded our analysis to cover the entire city, using data from all 147 sensors to evaluate the model's accuracy across different areas with varying pedestrian and car densities. This dual validation allows us to assess the model's performance across a broad range of conditions: high-density areas in the first experiment and widely varying densities in the second.} For the second experiment, we select 10 locations across Barcelona that represent diverse urban scenarios (see Fig.~ \ref{fig:manual_validation}b). The manual counts were acquired during two different periods on the same day: the first session from 11:00 a.m. to 1:00 p.m. and the second session from 5:00 p.m. to 7:00 p.m., both local time in Barcelona. \asr{These time frames approximately align with the periods of peak pedestrian activity on the streets, as shown in Fig.\ref{fig:exp_temporal}.} Each experiment involved manually counting the number of pedestrians passing the location over 5 minutes, followed by a 10-minute break, and concluding with another 5-minute counting session. The final count value for each location and period was obtained by averaging the results of the 2 sessions to obtain a more representative value of daytime pedestrian mobility volumes. 

Results of both experiments are reported in Table \ref{table:ped_flow_evaluation}. Overall, the error of the model is constrained, although the model tends to underestimate pedestrian counts. This general bias can be explained by the nature of the maximum entropy model used to estimate flows: in the absence of specific information, the algorithm distributes flows evenly between adjacent streets, while on some occasions, pedestrian traffic tends to be higher on certain pedestrian-friendly or commercial streets. In the case of manual counts, this bias is reduced and some sensors overestimate while others underestimate pedestrian volumes. Interestingly, manual counts achieve relatively lower errors compared to LOOCV results, despite being conducted in the 10 districts of Barcelona with a larger range of varying conditions.


\begin{table}[h]
\begin{center}
\begin{adjustbox}{max width=\textwidth}
\begin{tabular}{|l|c|c|c|c|c|c|c|c|c|c|c|}
\cline{2-12}
\multicolumn{1}{c|}{} & \multirow{2}{*}{\makecell{Num. \\ Experiments}} & \multicolumn{5}{|c|}{Absolute Error} & \multicolumn{5}{|c|}{Relative Error} \\
 \cline{3-12}
\multicolumn{1}{c|}{} &  & Mean & Std. & $Q_{25}$ & $Q_{50}$ & $Q_{75}$ & Mean & Std. & $Q_{25}$ & $Q_{50}$ & $Q_{75}$ \\
\hline
\hline
LOOCV & 75 & -580 & 473 & -843 & -563 & -246 & -0.653 & 0.374 & -0.957 & -0.847 & -0.380 \\
\hline
\makecell{ Manual street- \\ side counts} & 10 & -209 & 254 & -181 & -45 & 119 & -0.385 & 0.459 & -0.652 & -0.199 & 0.590 \\
\hline
\end{tabular}
\end{adjustbox}
\end{center}
\caption{\revone{Analysis of the accuracy of the pedestrian flow estimation over all sensors during 2019 at hourly resolution using Leave-One-Out Cross-Validation (LOOCV) and external manual counts. Absolute Error is displayed in Ped. h$^{-1}$. \textit{Std.} stands for standard deviation. $Q_{25}$, $Q_{50}$, $Q_{75}$ are the lower, median, and upper quartiles of the error, respectively.}}
\label{table:ped_flow_evaluation}
\end{table}

\subsection{High spatial-resolution NO$_2$ values}
\label{sec:methods_no2}
\begin{figure}[h!]
\begin{center}
    \includegraphics[width=.9\textwidth]{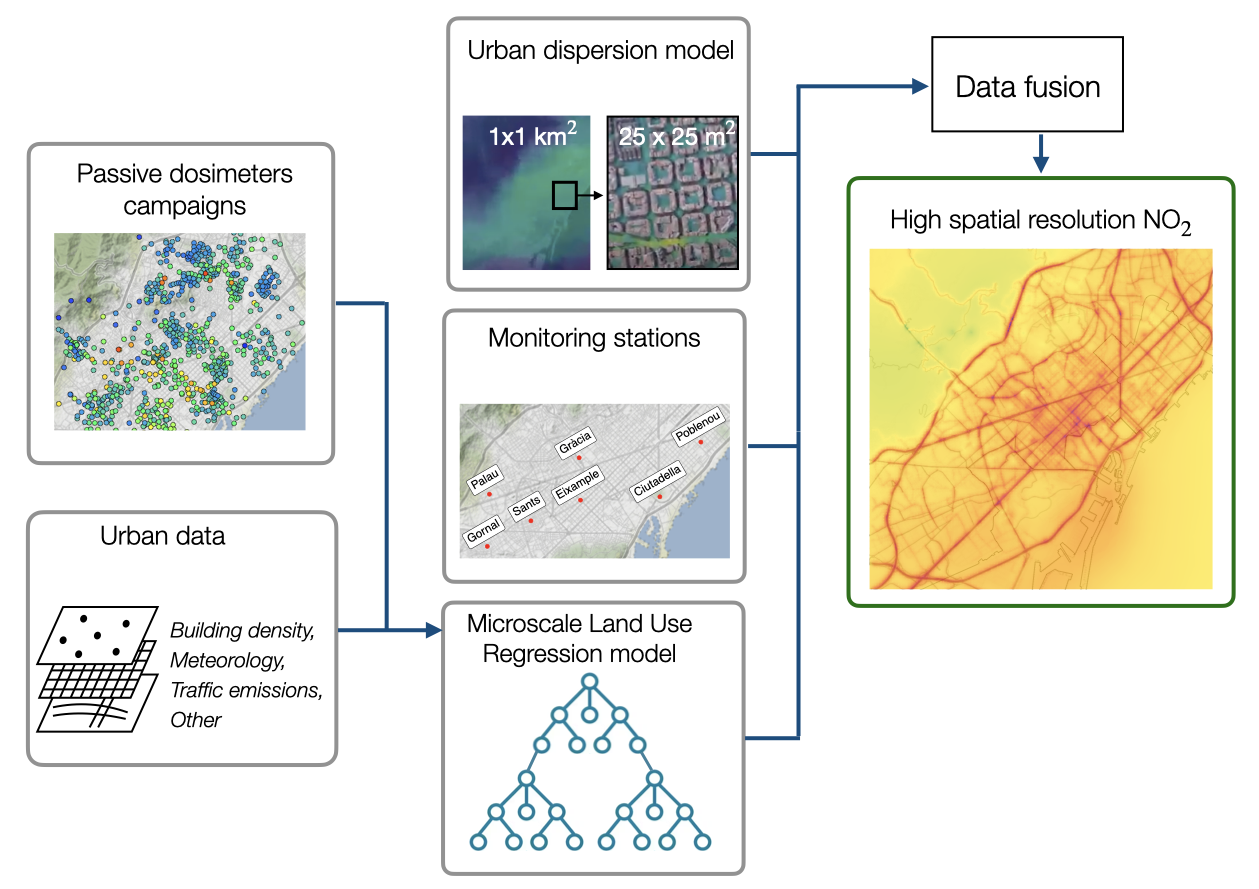}
  \end{center}
    \caption{Data-fusion pipeline for the estimation of high spatial resolution NO$_2$}
    \label{fig:pipeline_NO2}
\end{figure}

To align pedestrian mobility demand to hourly $NO_{2}$ data for the year 2019 we used a multi-scale high-spatial resolution urban air quality system \cite{benavides2019}, corrected with observational data coming from monitoring stations and passive dosimeters campaigns \cite{criado2023data}. 
This correction process consists in a data-fusion methodology based on Universal Kriging, which is a popular geostatistical technique \cite{cressie2015statistics}. Fig.~\ref{fig:pipeline_NO2} exhibits the main datasets and process employed to obtain the NO$_2$ data at the street scale. Our approach leverages hourly monitoring station data to correct the NO$_2$ short-term temporal behaviour, and short intensive passive dosimeter campaigns to adjust long-term spatial patterns at the street scale. \revone{The aforementioned Kriging-based} data-fusion process has already been validated for the domain and time period of the present study in \cite{criado2023data}. However, for the sake of completeness, we outline here the main components of this data-fusion process. 

\subsubsection{Urban dispersion model}
    The multi-scale air quality system CALIOPE-Urban \cite{benavides2019} produces hourly estimates of NO$_2$ surface concentrations over the Barcelona agglomeration. This system couples the regional air quality system CALIOPE \cite{baldasano2011annual} with the Gaussian dispersion model R-LINE \cite{snyder2013rline, venkatram2013re} adapted to street canyons. At the street scale, the chemical balance between NO$_x$ and NO$_2$ is simplified using the generic reaction set implemented in the R-LINE solver by Valencia \textit{et al.} \citep{valencia2018development}. Emissions are computed using the HERMESv3 bottom-up emission model \citep{guevara2020hermesv3}. The regional model CALIOPE, used to compute background concentrations, is based on a set of three nested domains (Europe at 12 km $\times$ 12 km, the Iberian Peninsula at 4 km $\times$ 4 km, and the region of Catalonia at 1 km $\times$ 1 km horizontal resolutions) \cite{Pay2014}. The urban model uses a road-following mesh, which refines the mesh up to  25 m $\times$ 25 m at regions of strong NO$_2$ gradients, and gradually degrades it until  500 m $\times$ 500 m resolution where gradients are low.

\subsubsection{Monitoring stations}
    Observations of hourly NO$_2$ surface concentrations are obtained from the official monitoring network in Catalonia \cite{XVPCA} (XVPCA) managed by the Catalan Regional Administration. There are 13 stations in the Barcelona agglomeration, from which 2 are traffic representatives, 3 are sub-urban background representatives, and 8 are urban background representatives. However, the data-fusion method excludes the Observatori Fabra station (sub-urban background representative) since it is the only one outside the urban canopy layer and exhibits different $NO_2$ dynamics, degrading the data-fusion skills within the urban canopy \cite{criado2023data}.
    
  \subsubsection{Microscale Land Use Regression (LUR) model based on passive dosimeter campaigns}
    Long-term spatial distribution of NO$_2$ is included in the data-fusion process using a machine-learning microscale LUR \cite{criado2023data} model, considered as a climatology of NO$_2$.  The target variable of the LUR model corresponds to observed $NO_2$ from 2 passive dosimeters campaigns \cite{xAire_data,benavides2019} (a total of 840 samplers distributed throughout the city), while the predictors (or features) comprise (i) building density \cite{institut.catalunya}, (ii) traffic intensity from the road-link traffic network of the HERMESv3 \cite{guevara2020hermesv3}, (iii) annual average data from the regional CALIOPE air quality system (NO$_2$ surface concentration, planetary boundary layer height, and wind speed from the lower model layer), and (iv) annual average NO$_2$ from the raw CALIOPE-Urban model. The LUR model relies on a Gradient Boosting Machine (GBM) algorithm to derive non-linear relations between predictors and the target variable. The selection of predictors, optimization of GBM hyperparameters, and the cross-validation evaluation are detailed elsewhere \cite{criado2023data}. The LUR code is available via Zenodo \cite{alvaro_criado_romero_2022_7185913}.

\subsection{Exposure metrics}
\label{sec:exp_metrics}
Throughout the paper, we use and compare several metrics of exposure. In this section, we incrementally introduce each metric and explain their relationships. The most simple and the first metric we define is the cumulative exposure $E_c$ which quantifies the total amount of a substance to which a given population is subjected over a specified period:
\begin{equation}
    {\langle E_{\mbox{\textsubscript{c}}}\rangle}_{[t_1,t_2]} = \int_{t_1}^{t_2} C(t) \cdot P(t) \ dt,
    \label{eq:PE_S}
\end{equation}
\noindent where $t_{1}$ and $t_{2}$ specify the time period, and $C(t)$ and $P(t)$ correspond respectively to the pollutant concentration and population over time in a given place. See that $E_I(t) = C(t) \cdot P(t)$ corresponds to the instantaneous exposure of the population at time $t$. The cumulative exposure, $E_c$, has been used in several works to quantify population exposure \cite{picornell2019population} and to identify exposure hot-spots \cite{santiago2021estimates}. From this original definition, it is useful to consider the average exposure a given population is subjected to over a specified period:
\begin{equation}
    {\langle{E}\rangle}_{[t_1,t_2]} = \frac{1}{t_2 - t_1} \int_{t_1}^{t_2} C(t) \cdot P(t) \, dt = {\langle C \rangle}_{[t_1,t_2]} \cdot {\langle P \rangle}_{[t_1,t_2]} + \text{cov}(C,P),
    \label{eq:PE_S_2}
\end{equation}

\noindent where angle brackets denote the average over the specified time window, and $\text{cov}(C,P)$ is the covariance between concentration and population presence. Note that if $\text{cov}(C,P)=0$, the computation simplifies significantly. This condition implies independence between $C$ and $P$, allowing for the use of individual measurements that are not temporally aligned---a potential advantage often overlooked in data analysis.

Exposure studies under the static approach typically use constant population estimates over census areas; and consequently, $\text{cov}(C,P) = 0$. In this case, the static exposure at census area $s$ is simply given by

\begin{equation}
    {\langle E \rangle}_{s,[t_1,t_2]} = {\langle C \rangle}_{s,[t_1,t_2]} \cdot P_s, ~~\mbox{with}~~{\langle C \rangle}_{s,[t_1,t_2]} = \frac{1}{N(t_2-t_1)}\sum_{n=1}^{N}\int_{t_1}^{t_2}C_{(x_n,y_n)}(t)dt,
\label{eq:static_exposure}
\end{equation}
\noindent where ${\langle C \rangle}$ corresponds to the time- and spatial-averaged concentration, $P_s$ the population in census area $s$, and  $(x_n,y_n)$ are the $N$ residential addresses within area $s$. This approach involves strong approximations as it explicitly assumes that residents stay permanently at home and are directly exposed to outdoor concentrations. In reality, indoor concentrations depend on both indoor sources (such as cooking or heating) and outdoor pollution that infiltrates the indoor environment \cite{hu2020relationship}. Despite these assumptions, this static approach is the most commonly used in epidemiological studies. It has shown to be sufficiently accurate to reveal statistically significant correlations between air quality and public health (e.g., \cite{pope2009fine}).

\asr{To map the pollution exposure risk of pedestrians, we propose a metric based on the rate of exposure, which accounts for both pedestrian flow $\left[\text{pedestrian}/\text{h} \right]$ and pollutant concentrations $\left[\mu \text{g} / \text{m}^3 \right]$. Unlike other measures that tend to focus on marginal or cumulative exposure over time or area, our approach offers a fundamental analysis that can later be extended to such metrics if needed, as we will then do.  Our \textbf{\textit{area-focused}} pedestrian exposure indicator quantifies the instantaneous exposure (exposure at a specific moment in time at a specific location) of the pedestrians, without assumption on their walking speed. The measure, $\langle E_{re} \rangle$, is therefore expressed in $\left[\mu \text{g} / \text{m}^3 \cdot \text{pedestrian}/\text{h} \right]$ units }and is specifically designed to quantity exposure of foot travelers using the pedestrian infrastructure. Conveniently, this indicator is independent of sidewalk length, allowing us to identify exposure hot-spots even on very short sidewalks.

Defining $F(l,t)$ as the pedestrian flow intensity on sidewalk $l$ at time $t$, and $C(l,t)$ the corresponding concentration of $NO_2$; the Pedestrian Average Rate of Exposure in area $a$ over a time interval $[t_1,t_2]$ can be expressed as:
\begin{align}
{\langle E_{\mbox{\textsubscript{re}}}\rangle}_{L_a,[t_1,t_2]}& = \frac{1}{|L_a|(t_2-t_1)}\sum\limits_{i=1}^{|L_a|} \int_{t_1}^{t_2} C(l_i,t) \cdot F(l_i,t) dt \label{eq:exposure11}\\
&= \langle C\rangle_{L_a,[t_1,t_2]} \cdot \langle F\rangle_{L_a,[t_1,t_2]} + \mbox{cov}(C,F),
\label{eq:PE_S_3}
\end{align} 

\noindent where $L_a$ is the set of all sidewalks within area $a$, bars indicate the cardinality of the set, and $\langle C \rangle$ and $\langle F \rangle$ are the time-averaged concentration level and pedestrian flow associated at sidewalk segment in area $L_a$. Note that Eq. \ref{eq:PE_S} can also be applied to a single sidewalk if needed. In that case, with a slight abuse of nomenclature, we will denote it as $\langle E_{\text{re}}\rangle_{l_i,\circ}$. See also that if we have several measurements per sidewalk, one may need to either average these measurements or incorporate them into the model for a finer level of analysis. In our dataset, we primarily have one measurement per sidewalk segment, so this serves as our approximation for exposure at that specific segment.

Finally, to geographically compare the static exposure at the census level, $\langle E\rangle_{s,\circ}$, with the exposure experienced by pedestrians on sidewalks, $\langle E_{\text{re}}\rangle_{l_i,\circ}$, we propose to aggregate the rate of exposure along sidewalk segments at census area scale. Thus, considering the length of each sidewalk segment $\left\Vert l_i \right\Vert$ and an average speed of pedestrian $V_{walk} = 1.34 \ m/s$ \cite{bosina2017estimating,rhoads2023inclusive}, pedestrian exposure at census area $s$ is obtained as
\begin{equation}
    \langle E{\mbox{\textsubscript{ped}}} \rangle_{s,[t_1,t_2]} = \sum_{i=1}^{|L_s|} \langle E_{\text{re}}\rangle_{l_i,[t_1,t_2]} \cdot \frac{\left\Vert l_i \right\Vert}{V_{walk}},
    \label{eq:PPE}
\end{equation}

\noindent where $|L_s|$ is the total amount of sidewalks segments in census area $s$. Table \ref{table:indicators} summarizes all exposure indicators used in the present study.

\renewcommand{\arraystretch}{1.3}
\begin{table}[h]
\begin{adjustbox}{max width=\textwidth}
\begin{tabular}{|l|c|c|c|}
\hline
\textbf{Indicator}   & \textbf{Scale} & \textbf{Definition } \\
\hline \hline
Static Exposure at census area $s$ &  Census area & $\langle E \rangle_{s,\circ}$ \\
\hline
Pedestrian Average Rate of Exposure at sidewalk $l$&  Sidewalk/Crossing &  $\langle{E_{\mbox{\textsubscript{re}}}}\rangle_{l,\circ}$ \\
\hline
Pedestrian Population Exposure at census area $s$ &  Census area & $\langle {E{\mbox{\textsubscript{ped}}}} \rangle_{s,\circ}$  \\ 
\hline
\end{tabular}
\end{adjustbox}
\caption{The different metrics for exposure indicators used in this study. Scale refers to spatial aggregation. }
\label{table:indicators}
\end{table}
\renewcommand{\arraystretch}{1}

\section{Results}

This section is divided into two parts. Subsection \ref{sec:temporal} examines the temporal profiles of NO$_2$ values, pedestrian flows, and exposure at the pedestrian count sensors, providing results for a scenario where only NO$_2$ levels have been estimated. Subsection \ref{sec:spatial} presents a city-scale analysis of pedestrian exposure, focusing on spatial aggregations at sidewalks and census areas.

\subsection{Temporal profiles at pedestrian counter locations}
\label{sec:temporal}

\begin{figure}[h!]
    \includegraphics[width=\textwidth]{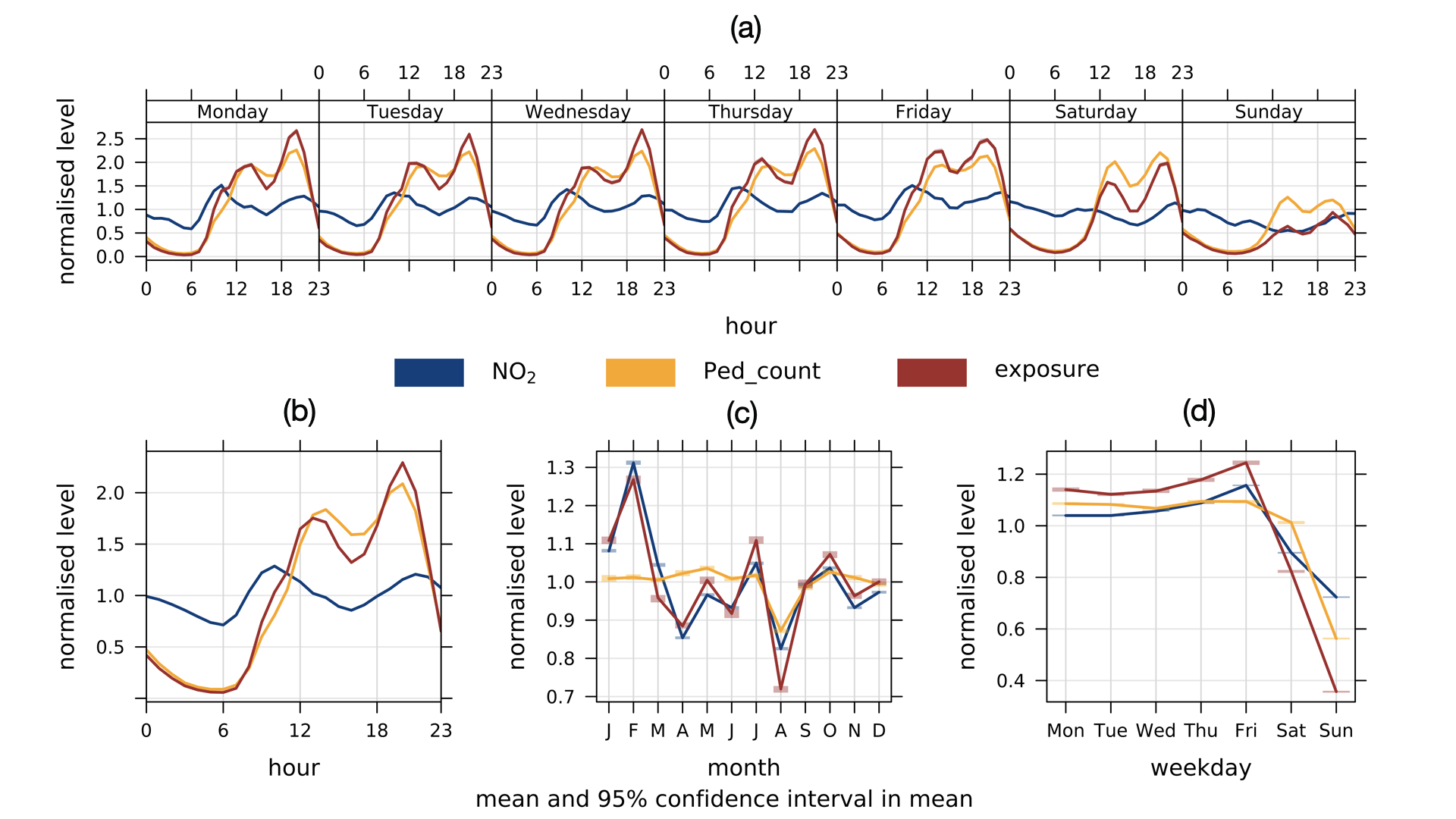}
    \centering
    \caption{ \asr{NO\textsubscript{2} concentration, pedestrian flow, and exposure profiles normalized by their respective averages at pedestrian counters. Results are computed by averaging all hourly sensor data in 2019. (a) Daily profiles for each day of the week, (b) averaged daily profiles, (c) annual profiles, and (d) weekly profiles. For the case of annual and weekly profiles, the 95\% confidence intervals in the mean can be seen.}}
    \label{fig:exp_temporal}
\end{figure}

As shown in Eq.~\ref{eq:PE_S}, pedestrian exposure at a specific location depends on the temporal alignment of pedestrian flow intensities and NO$_2$ concentrations.  Thus, in the following subsections, we analyze the temporal profiles of these variables and explore their potential correlation. The temporal profiles of the rate of exposure, obtained with Eq.~\ref{eq:exposure11}, are shown in Fig. \ref{fig:exp_temporal}, together with pedestrian flows and NO$_2$ data. For each measure, the results are averaged over all sensor sites. To facilitate the comparison, we normalize all variables by their respective means. 

\subsubsection{Observed pedestrian counts}
The yellow lines in Fig.~\ref{fig:exp_temporal} display the measured hourly pedestrian flows aggregated at different time frames. At hourly resolution (Fig.~\ref{fig:exp_temporal}b), we observe a bimodal distribution with a slightly higher peak in the evening (around 20h UTC), a minor peak in the afternoon (around 14h UTC), and a sustained high pedestrian presence during daytime. At daily resolution, the behavior remains quite stable, with lower rates on weekends, especially on Sundays, as shown in  Fig.~\ref{fig:exp_temporal}d. The annual profile, shown in Fig.~\ref{fig:exp_temporal}c, indicates a noticeable drop in pedestrian counts during August. However, overall results suggest similar pedestrian counts throughout the year. A significant variability of pedestrian flows among sensors is detected: 25 \% of sensors have hourly peaks greater than 1500 pedestrian/h; while another 25  \% have peaks lower than 400 pedestrian/h (see Fig.~\ref{fig:ped_temporal}). These indicate a wide range of monitored sidewalks, from crowded pedestrian avenues to secondary streets. For further details on the absolute values of pedestrian flow over time, we refer the reader to Fig.~\ref{fig:ped_temporal} in Appendix A.

\subsubsection{Modelled NO$_2$ at pedestrian counter locations}

To provide a detailed analysis, we leverage fine-grain NO$_2$ estimations near traffic roads (25 m $\times$ 25 m grid resolution) from \cite{criado2023data} to interpolate NO$_2$ concentrations at the locations of the pedestrian counters. Dark blue lines in Fig.~\ref{fig:exp_temporal} show the NO$_2$ temporal profiles in 2019. As for pedestrian flow, the daily profile for NO$_2$ is consistent during the week (see Fig.~\ref{fig:exp_temporal}a), displaying a bimodal shape with a slightly higher peak in the morning (Fig.~\ref{fig:exp_temporal}b). The morning and evening peaks occur around 9 and 22h UTC, respectively, and correspond to traffic rush hours and adverse meteorological conditions for dispersion of pollutants, i.e., low to moderate values of the planetary boundary layer height \revtwo{($PBL \lesssim 500$ m)}. 
At monthly resolution (Fig.~\ref{fig:exp_temporal}c), there is a noticeable NO$_2$ peak around February and a sudden drop during August. Overall, in relation to the pedestrian densities at the same locations, we observe that the timing of NO$_2$ peaks differs from the observed pedestrian flow peaks.
Fig.~\ref{fig:NO2_temporal} in Appendix B shows the absolute values of the NO$_2$ temporal patterns. They reflect the wide range of NO$_2$ levels we detect, from highly trafficked ($>$100 $\mu g/m^3$) to traffic-calmed streets ($<$20 $\mu g/m^3$). Additionally, the relative variability of NO$_2$ data is considerably lower than that of pedestrian flows shown in Fig.~\ref{fig:ped_temporal}.

\revone{We also highlight that in 2019, several locations were observed where NO$_2$ values exceeded the current annual and hourly limits of 40 and 200 $\mu \text{g}/\text{m}^3$, respectively, set by the by the Directive 2008/50/EC on ambient air quality and cleaner air for Europe \cite{union2008directive}.} Particularly, several NO$_2$ hot-spots stand out on major trafficked streets, ring roads, the central district \textit{Eixample}, and the areas surrounding the port (see Fig.~\ref{fig:NO2_2019} in Appendix B for further details).   

\subsubsection{Estimated exposure at pedestrian counter locations}

A major insight of the previous analysis is that the daily profile of the rate of exposure (Fig.~\ref{fig:exp_temporal}b) is primarily driven by pedestrian flows rather than by temporal pollutant distribution. On the contrary, the monthly profile of pedestrian exposure is predominantly dominated by NO$_2$ levels (see Fig.~\ref{fig:exp_temporal}c). This indicates that short-term exposure (hourly resolution) is mainly explained by pedestrian flow intensity, while long-term exposure (monthly resolution) correlates with NO$_2$ temporal patterns. We can already conclude that these findings may have significant methodological implications on urban policy design. Depending on the targeted temporal scale of exposure, mitigation exposure policies should focus either on pedestrian rush hours or on synoptic time scales when pollution episodes occur.

\begin{figure*}[h]	
  \begin{center}
    \includegraphics[width=0.5\textwidth]{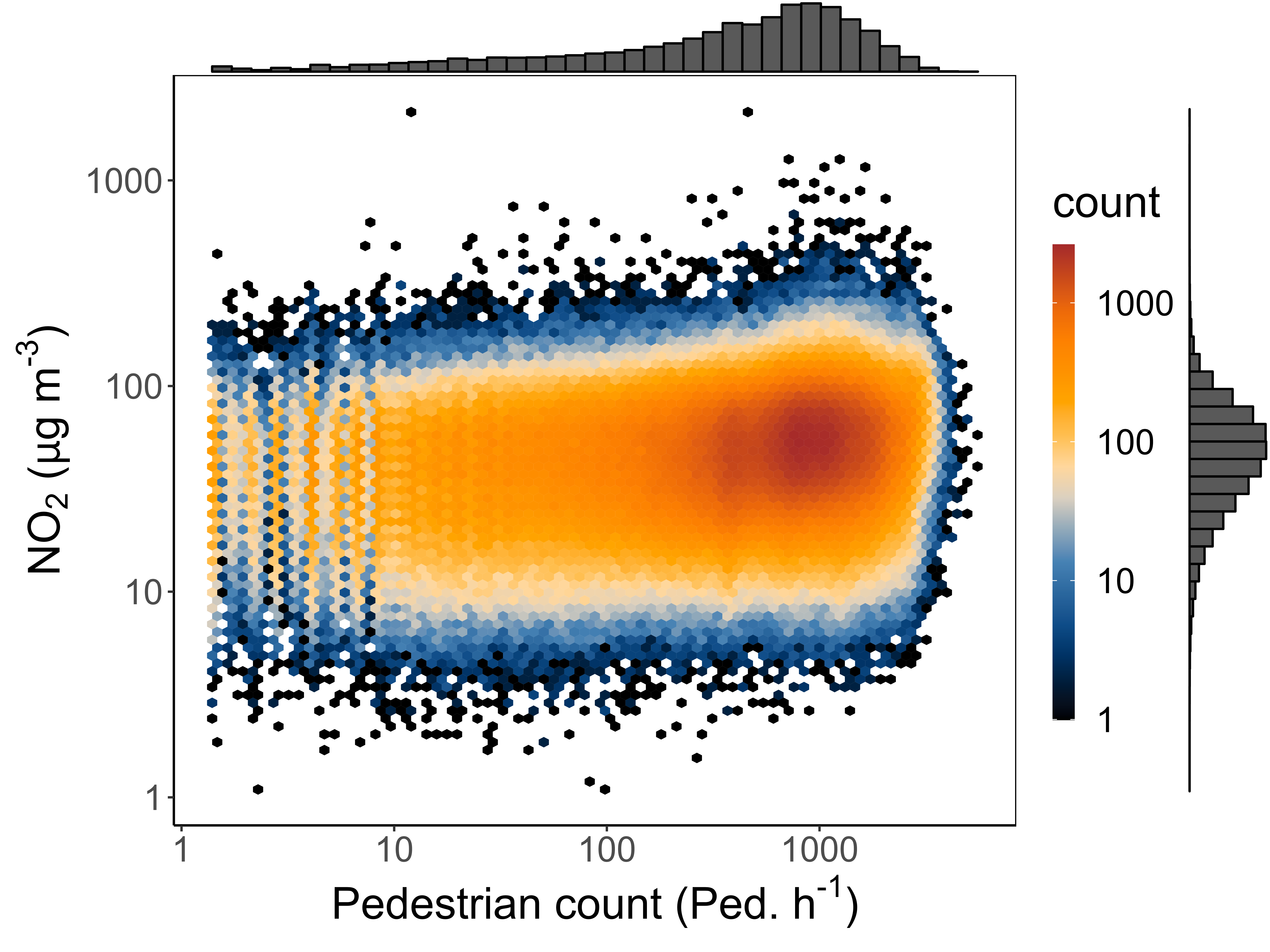}
  \end{center}
    \caption{(b) Relationship between binned counts of  pedestrian flow and NO\textsubscript{2} hourly values. The marginal distributions of each variable are shown above (Pedestrian count) and on the right side (NO\textsubscript{2}). Results encompass hourly data spanning the entirety of 2019.}
    \label{fig:exposure_temporal}
\end{figure*}

To explore the potential temporal correlation between NO$_2$ and pedestrian flows, Fig.~\ref{fig:exposure_temporal} shows the joint distribution of pedestrian intensities at sidewalks and their NO$_2$ concentrations, as well as their marginal distributions, which seem to follow long tail patterns. Interestingly, pedestrian and NO$_2$ practically do not show any correlation. To further analyse the potential correlations, Table \ref{table:UK_stats} quantifies ${\langle E_{\mbox{\textsubscript{re}}}\rangle}_{L,[2019]}$ ,$\langle C\rangle_{L,[2019]}$, $\langle F\rangle_{L,[2019]}$, and $\mbox{cov}(C,F)$ spatially aggregated at each cluster as defined in Fig.~\ref{fig:manual_validation}. The NO$_2$ concentration, pedestrian flows, and exposure values are averaged over each cluster for each hour. The covariance between NO$_2$ and pedestrian flow is then computed over the entire time series and aggregated by cluster area.
These results further confirm a minimal correlation between NO$_2$ and pedestrian flows and indicate that neglecting covariance implies a relative error of less than 10\%. Therefore, in the following sections, the spatial patterns of pedestrian exposure have been estimated as ${\langle E_{\mbox{\textsubscript{re}}}\rangle}_{L_a,[t_1,t_2]} \approx \langle C\rangle_{L_a,[t_1,t_2]} \cdot \langle F\rangle_{L_a,[t_1,t_2]}$. This approximation is highly advantageous in terms of data acquisition protocols and computational efficiency, as it allows to estimate exposure solely considering individual pedestrian flow and NO$_2$ averages.

\begin{table}[!h]
    \centering
    \begin{adjustbox}{max width=\textwidth}
    \begin{tabular}{ccccc}
        \textbf{Cluster name} &
        \makecell{ NO$_2$,  $\langle C\rangle_{L,[2019]}$ \\ $\left[\mu g / m^3 \right]$} &
        \makecell{ Ped. flow, $\langle F\rangle_{L,[2019]}$ \\ $\left[\text{ped.}/h \right]$} &
        \makecell{ Ped. Exp., ${\langle E_{\mbox{\textsubscript{re}}}\rangle}_{L,[2019]}$ \\ $\left[\mu g / m^3 \cdot \text{ped.}/h \right]$} &
        \makecell{ $\mbox{cov}(C,F)$ \\ $\left[\mu g / m^3 \cdot \text{ped.}/h \right]$} \\ \hline 
        Ciutat Vella      & 47.5 & 1167 & 58428 & 3021 \\ 
        Sants   & 52.4 & 948 & 53307 & 3586 \\ 
        Gràcia  & 57.2 & 719 & 45957 & 4839 \\ 
        Sant Antoni & 48.5 & 784 & 45933 & 7953 \\ 
        Eixample & 56.2 & 736 & 43356 & 1957 \\ 
        Les Corts & 46.5 & 808 & 37328 & -284 \\ 
        Sagrada Familia & 56.9 & 510 & 32176 & 3176 \\ 
        Sant Andreu & 38.7 & 493 & 20094 & 1003 \\ 
        Nou Barris & 38.9 & 471 & 19750 & 1432 \\ 
        Outliers (rest of the city) & 40.3 & 548 & 26586 & 4511 \\ \hline
    \end{tabular}
    \end{adjustbox}
    \caption{Summary statistics averaged over time and clusters. $\langle C\rangle$ represents the NO$_2$ concentration level, $\langle F\rangle$ corresponds to pedestrian flows, ${\langle E_{\mbox{\textsubscript{re}}}\rangle}$ denotes the pedestrian rate of exposure, and $\mbox{cov}(C,F)$ indicates the covariance between NO$_2$ concentration and pedestrian flow.}
    \label{table:UK_stats}
\end{table}

\subsection{Modelling exposure at a city-wide scale}
\label{sec:spatial}
After determining the temporal profiles of pedestrian exposure using sensor data, we proceed to examine its spatial patterns across Barcelona at different scales: sidewalks and census areas.

\subsubsection{Annual average pedestrian exposure on sidewalks}
\label{sec:results_exposure_sidewalks}

\begin{figure*}[h]	
    \includegraphics[width=\textwidth]{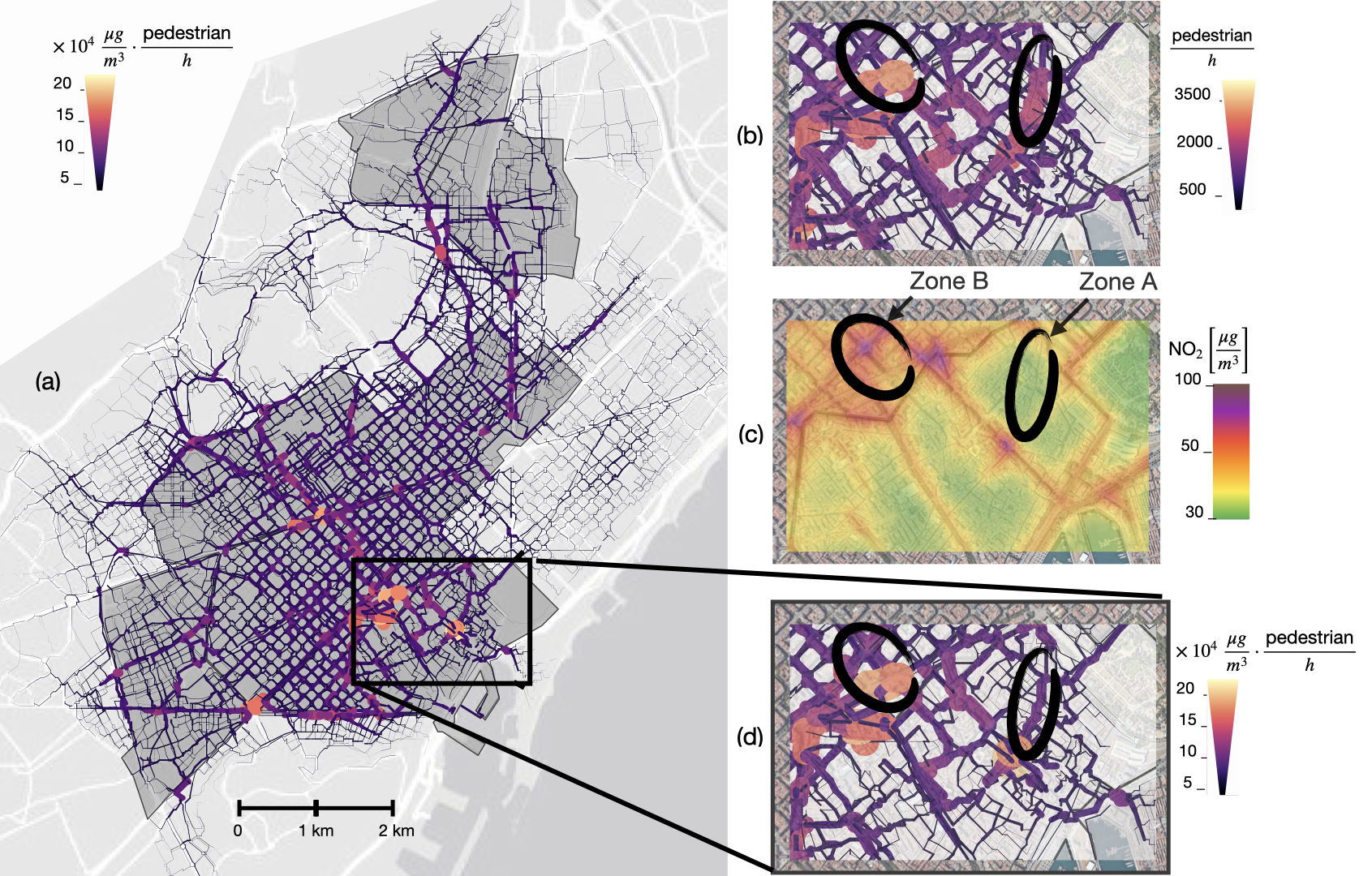}
    \caption{(a) Average pedestrian exposure across Barcelona for 2019, combining the time-averaged NO\textsubscript{2} levels interpolated at the center point of each sidewalk segment with the specific pedestrian flow intensity. The grey polygons in the background correspond to city areas where sensors are located, the definition of these clusters is detailed in section \ref{sec:ped_estimates}. Right panels (b-c) show a zoom-in of the city center to illustrate the annual averages of (b) pedestrian flows, (c) NO$_2$ surface concentration levels, and (d) pedestrian exposure on sidewalks.}
    \label{fig:sidewalks_results}
\end{figure*}

Fig.~\ref{fig:sidewalks_results}a shows the Pedestrian Average Rate of Exposure ${\langle E_{\mbox{\textsubscript{re}}}\rangle}$ to NO$_2$ across Barcelona city for 2019. In the figure, the different clusters of analysis, where most of the sensors are located, are shaded in gray for informative purposes. At a global scale, the results show high pedestrian exposure in the city center, particularly at the intersection of the \textit{Eixample}, \textit{Sant Antoni}, and \textit{Ciutat Vella} clusters (see Fig.~\ref{fig:manual_validation} for clusters names). While this is an expected result, as city centers typically accumulate high vehicle and pedestrian traffic, strong gradients exist within these areas when examined closely (Fig.~\ref{fig:sidewalks_results}d). This underscores the importance of using high-resolution spatial data to accurately capture pedestrian exposure and inform public actions aimed at minimizing pedestrian exposure risk. As demonstrated in our analysis, the level of detail required to accurately quantify exposure in urban environments is critical, given that NO$_2$ levels and pedestrian flows can vary significantly even between closely situated regions.

Figs.~\ref{fig:sidewalks_results}b, \ref{fig:sidewalks_results}c, and \ref{fig:sidewalks_results}d detail the regions where most of the exposure hot-spots are located. \revtwo{These panels illustrate salient interactions among pedestrian flow intensities, NO$_2$ concentration levels, and pedestrian exposure}. In the panels, we highlight two different regions. Zone A corresponds to a traffic-calmed street that has high values of pedestrian flows (around 2200 pedestrian/h) and relatively low NO$_2$ concentration levels (around 30 $\mu g/m^3$ ). Zone B has lower values of pedestrian flow but extremely poor air quality conditions (NO$_2$ concentration around 80 $\mu g/m^3$). Accordingly, exposure is lower in zone A (around 6.6 $\times 10^4$ $\mu g/m^3 \cdot$ pedestrian/h) and higher in zone B (around 15 $\times 10^4$ $\mu g/m^3 \cdot$ pedestrian/h), despite these two regions being very close each other. This illustrates that not all areas with large pedestrian flows directly translate to hot-spots of exposure. Despite exposure being strongly dominated by pedestrian flows, NO$_2$ plays an important role by diminishing or raising exposure on crowded sidewalks.

\subsubsection{Residents' versus pedestrians' exposure}
\label{sec:results_exposure_areas}

\begin{figure*}[!h]
	\begin{tabular}{ll}		
        (a) & (b) \\	
        \includegraphics[width=.47\textwidth]{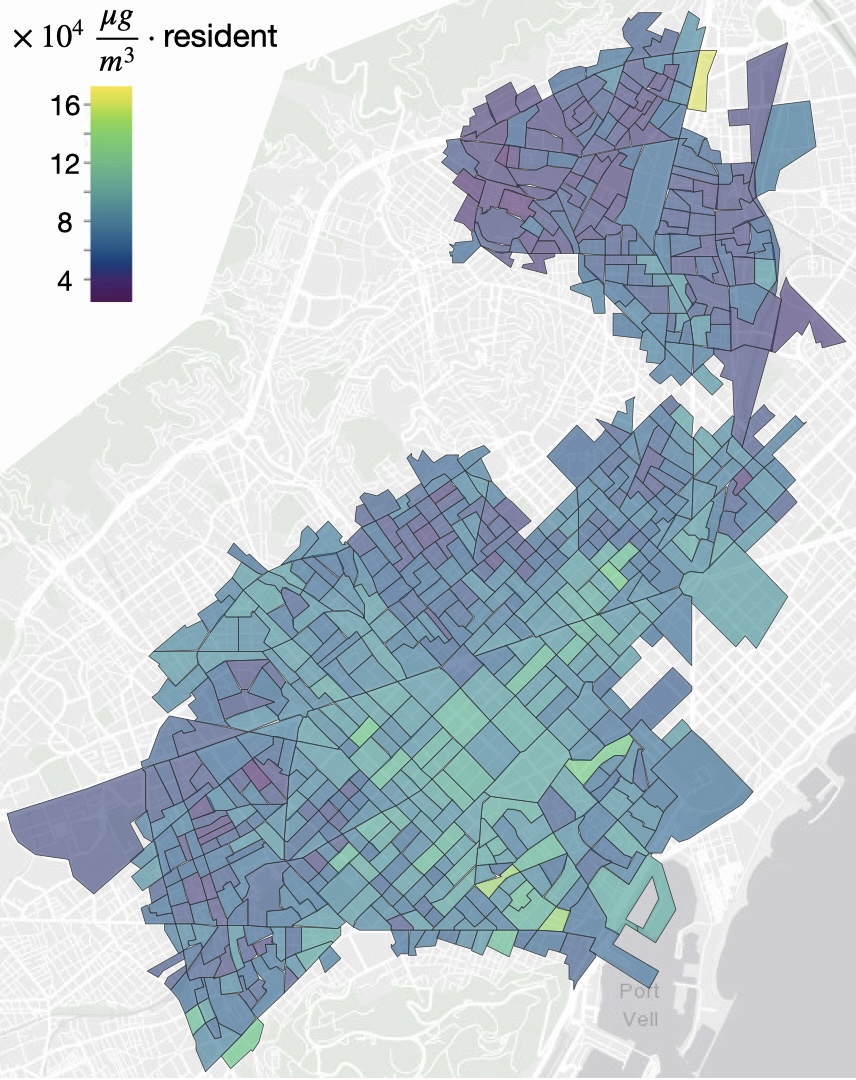} 
        &
        \includegraphics[width=.47\textwidth]{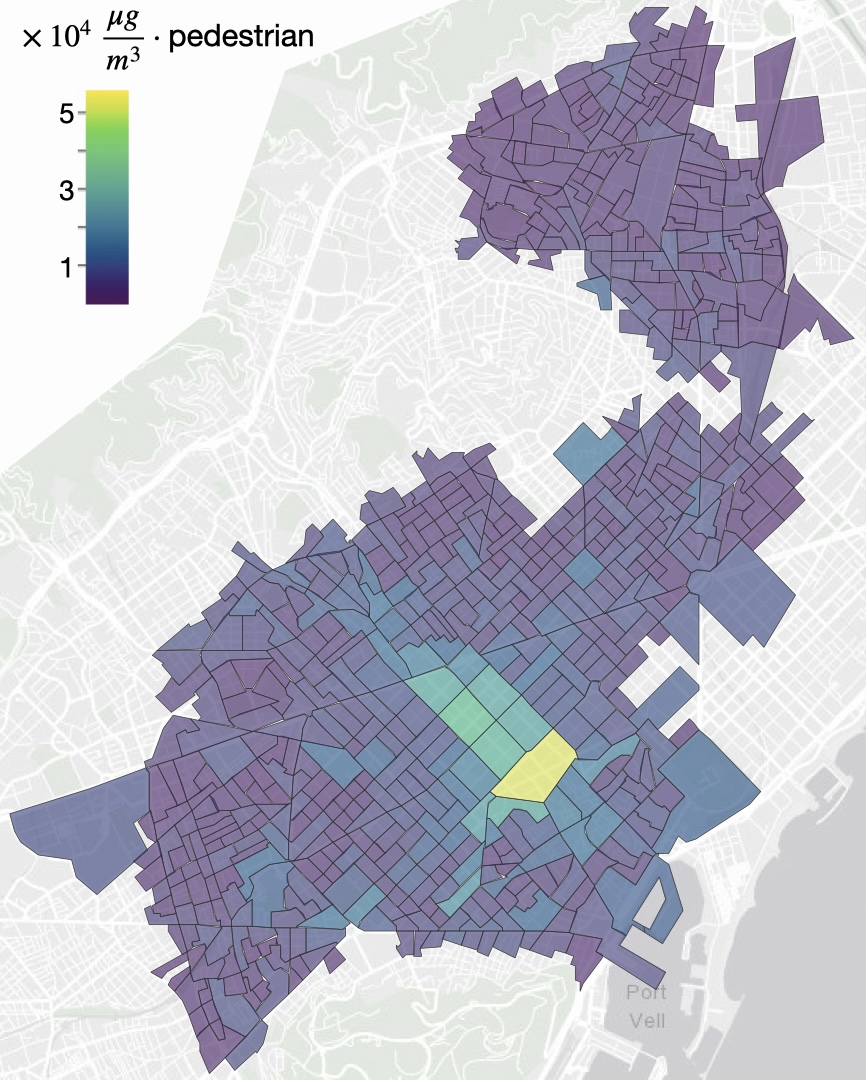}
    \end{tabular}
    \caption{(a) Static resident exposure as the product of each area's population by the annual averaged NO$_2$ weighted by residential addresses (Eq.~\ref{eq:static_exposure}). (b) corresponds to the pedestrian exposure at the census area (Eq.~\ref{eq:PPE}).}
    \label{fig:ped_vs_resident_exp}
\end{figure*}

As justified above, a common measure of exposure is to consider static approaches with a constant census population \cite{pope2009fine}. However, when accounting for population mobility, other metrics arise that may differ from the static approach. Following the exposure metrics detailed in Sec.~\ref{sec:exp_metrics}, we address here this problem and compare the annual average pedestrian exposure with the static census exposure for 2019. Results are shown in Fig.~\ref{fig:ped_vs_resident_exp} for all census areas intersecting the sensor clusters presented in Fig.~\ref{fig:manual_validation}. Exposure intensities are markedly higher for residential exposure due to assumptions of the static approach, which inaccurately considers that residents constantly stay indoors and directly inhale outdoor air. This approximation fails to consider varying daily activities and their environmental interactions. Additionally, proximity to pollution sources significantly impacts health, which leads us to study pedestrian exposure due to their frequent closeness to traffic. Disregarding the magnitude of the measure, we would like to emphasize the pronounced differences in exposure's spatial distribution across different citizen subgroups. Residential exposure is roughly consistent citywide, whereas pedestrian exposure is concentrated in specific hot-spots. These disparities indicate that urban policies to reduce exposure should be tailored to specific citizen subgroups. For instance, initiatives aimed at reducing residential exposure could include city-wide measures like Low-Emission Zones or urban tolls. Conversely, strategies targeting pedestrian exposure should concentrate on local hot-spots by altering pedestrian flows or implementing local measures such as tactical urban planning or \textit{superblocks} (or traffic-calmed streets) to separate highly trafficked streets from pedestrian areas.

\section{Conclusions}


\revtwo{
In this study, we developed a methodology to assess city-wide pedestrian exposure using {\it area-focused} indicators tailored to urban planning rather than epidemiological analysis.

Our findings reveal that daily pedestrian exposure profiles (at hourly resolution) are predominantly influenced by pedestrian flows, as pedestrian traffic shows high variability compared to relatively stable NO$_2$ levels. Consequently, strategies to mitigate exposure peaks should focus on managing pedestrian flows in high-risk areas. This can be achieved by reducing vehicle traffic, rerouting pedestrian pathways, or introducing physical barriers between roads and sidewalks. Traffic control measures targeting peak pedestrian hours (14:00 and 20:00 UTC) could further reduce NO$_2$ levels during these critical periods.

Conversely, pedestrian exposure at the monthly resolution is driven largely by fluctuations in NO$_2$ concentrations, often linked to pollution episodes that persist for several days. Therefore, mitigating long-term exposure peaks requires strategies that address broader temporal trends in air quality.

Given the pronounced spatial gradients in exposure, we emphasize the importance of combining detailed pedestrian mobility models with high-resolution pollution data to accurately assess pedestrian exposure. A city-wide evaluation of sidewalk infrastructure enabled us to identify highly localized exposure hotspots. This suggests that localized interventions could be highly effective.

At the census level, residential exposure tends to be uniform across the city, while pedestrian exposure is concentrated in the city center and certain hotspots. These distinct patterns imply that mitigation strategies should be tailored to the specific needs of different citizen groups.

The {\it area-focused} exposure indicator captures the unique spatiotemporal patterns of pedestrian exposure. Urban planners and policymakers can use this insight to design effective strategies that reduce exposure without diminishing the benefits of active mobility, ensuring vibrant, healthy urban environments while protecting the public from harmful pollutants.
}

\section*{Funding}
JM. A., C.C., A.C., and A.S. acknowledge financial support from the AIR-URBAN project (TED2021-130210A-I00/ AEI/10.13039/501100011033/ European Union NextGenerationEU/PRTR). A.S.-R., and J.B.-H. acknowledge financial support from the Ministry of Science and Innovation of Spain, through project No. PID2021-128966NB-I00, and (with C.R) from the Ajuntament de Barcelona and Fundació La Caixa (Spain), Project No. 21S09383-001. J.B.-H. acknowledges financial support from the Ramón y Cajal program through the grant RYC2020-030609-I. 

\setcounter{figure}{0}
\setcounter{table}{0}

\appendix

\section{A Simple Gravity Model for Baseline Mobility Estimation in Barcelona} \label{sec:OD}

We construct our initial mobility estimate for Barcelona using a gravity model (Block 2 in Fig.~\ref{fig:ped_estim_pipeline}). The resulting mobility demand is represented as an OD matrix, denoted by $o_{ij}$ (refer to Block 3 in Fig.~\ref{fig:ped_estim_pipeline}). The OD-Matrix represents the hourly volume of pedestrians departing from location $i$ with target destination $j$. Although this data is not strictly required for the estimation, it facilitates convergence, given the heterogeneous distribution of sensors across the various districts of the city. For this initial estimation, we rely on a gravity model implemented as follows. We consider a combination of population data, public transit stops and Points of Interest (POIs), which comprise all shops and public services in Barcelona. First, we calculate the node-level masses, $w_i$, for each network node as a weighted sum of local attributes: 
\begin{equation}
    w_i = \gamma_p \mbox{population} + \gamma_o \mbox{numPOIs} + \gamma_b \mbox{busStops} + \gamma_m\mbox{metroStops}
    \label{scaling_function_OD}
\end{equation}
The values of \textit{population} represent the adjacent housing block-level population data; \textit{numPOIs}, \textit{busStops} and \textit{metroStops} respectively denote the number of POIs, bus and metro stations adjacent to the edges connected to that node. Weighting factors, $\gamma_{\circ}$, help to calibrate the relative importance of these various spatial urban characteristics influencing pedestrian mobility. The distance decay (probability of a trip between $i$ and $j$ as a function of their distance) and the physiological limitations of pedestrian mobility are accounted for by incorporating a half-normal function with a standard deviation of 750 meters. This ensures that 95\% of the trips remain within a 15-minute walking distance. \asr{This value is more conservative than the average walking trip duration for residents of Barcelona (approximately 14 minutes) \cite{emef}, but statistics notably excludes a portion of the floating population, such as tourists.} The resulting OD matrix is therefore computed considering 
\begin{equation}
    \label{eq:gravity}
    o_{ij} =\alpha~w_i~w_j'~e^{-d_{ij}^2/2\sigma^2}
\end{equation}
where $d_{ij}$ is the network distance between any node $i$ and $j$, $w_i$ the vector of weights at node $i$ computed as above, and $\alpha$ is a scaling factor. The different parameters of the model are calibrated as follows. \asr{The weighting factors were manually optimized based on the correlation between the counts generated by the gravity model (in terms of $R^2$) and the sensor data, assuming shortest-path routing. The resulting values are $\gamma_p=1$, $\gamma_o=4$, $\gamma_b=100$ and $\gamma_m=160$.} Finally, the total number of trips in the city was calibrated using $\alpha$ to align with data from the Barcelona Daily Mobility Survey (EMEF) \cite{emef}, which recorded 3 million daily walking trips within Barcelona in 2019.

\section{Pedestrian count temporal variability}
\label{sec:sample:appendix}
\begin{figure}[h!]
    \includegraphics[width=\textwidth]{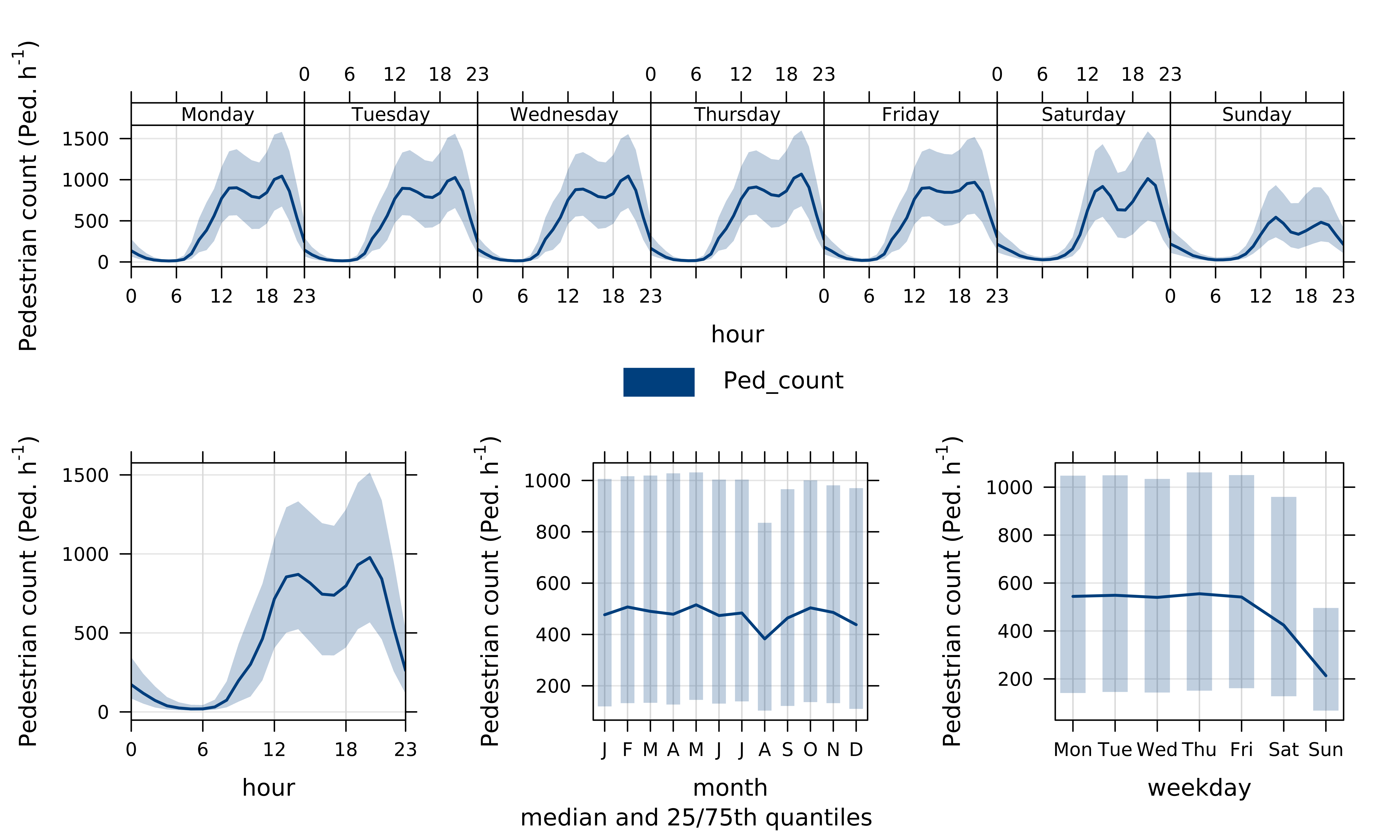}
    \centering
    \caption{Pedestrian count median (solid line) and 25 and 75th quantiles (shaded) of all sensors. Hourly, weekly, and monthly medians are represented. Observations have hourly resolution and they are averaged over 193 available sensors deployed over Barcelona city during 2019. }
    \label{fig:ped_temporal}
\end{figure}
\section{NO$_2$ count temporal variability}

\begin{figure}[t]
    \includegraphics[width=0.5\textwidth]{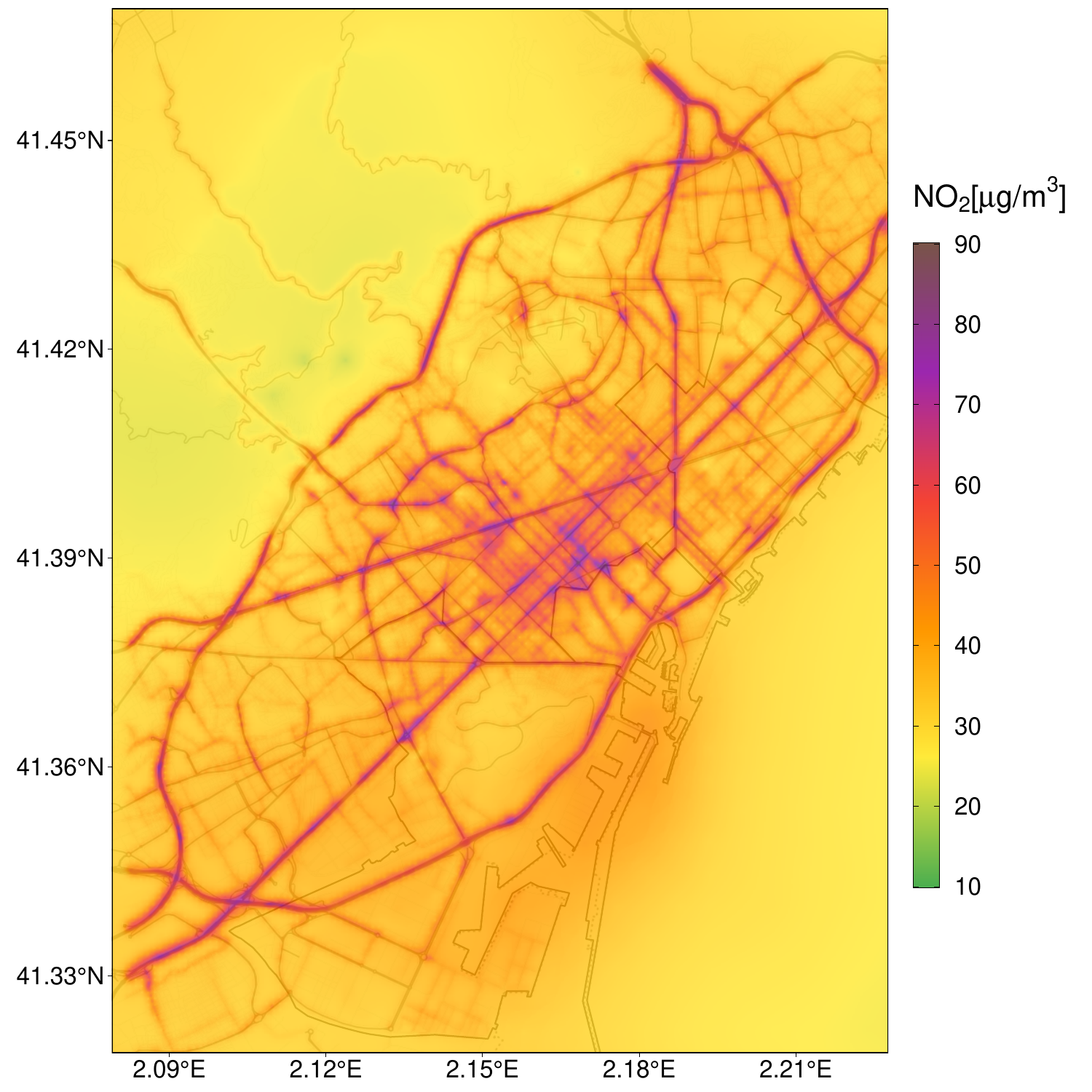}
    \caption{Annual average of NO\textsubscript{2} surface concentrations in 2019 in Barcelona.}
   \label{fig:NO2_2019}
\end{figure}

\begin{figure}[h!]
    \includegraphics[width=\textwidth]{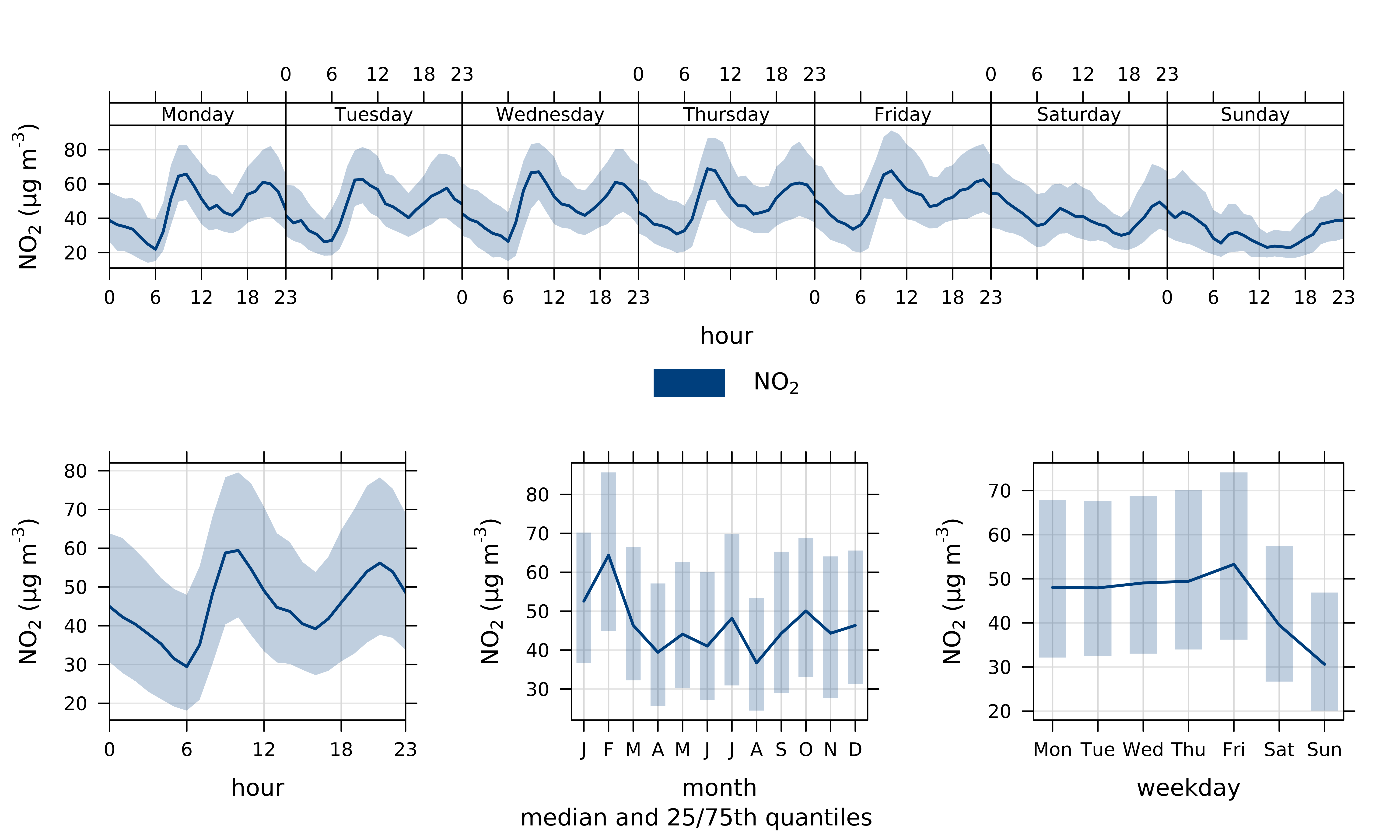}
    \centering
    \caption{NO\textsubscript{2} median (solid line), and 25 and 75th quantiles (shaded) of model data interpolated at all sensors locations. Hourly, weekly, and monthly averages are represented. Again, data have hourly resolution and have been averaged over the 193 sensors deployed over Barcelona during 2019.  }
    \label{fig:NO2_temporal}
\end{figure}



\end{document}